\documentclass[twocolumn,showpacs,preprintnumbers,amsmath,amssymb,prb,floatfix]{revtex4}

\usepackage{epsfig,psfrag}
\usepackage{color}
\usepackage{graphicx}% Include figure files
\usepackage{dcolumn}% Align table columns on decimal point
\usepackage{bm}% bold math

\begin{document}

\title{Josephson effect for ${\bf SU(4)}$ carbon nanotube quantum dots}

\author{A. Zazunov,$^{1}$ A. Levy Yeyati,$^2$ and R. Egger$^1$}
\affiliation{
$^{1}$Institut f\"ur Theoretische Physik, Heinrich-Heine-Universit\"at, 
D-40225 D\"usseldorf, Germany \\
$^2$Departamento de F{\'i}sica Te{\'o}rica de la Materia Condensada C-V,
Universidad Aut{\'o}noma de Madrid, E-28049 Madrid, Spain
}
\date{\today}
\begin{abstract}
We present the theory of the Josephson effect in nanotube dots
where an $SU(4)$ symmetry can be realized.
We find a remarkably rich phase diagram that significantly differs
from the $SU(2)$ case. In particular, $\pi$-junction behavior
is largely suppressed.  We analytically obtain the Josephson 
current in various parameter regions: (i) in the Kondo regime, 
covering the full crossover from $SU(4)$ to $SU(2)$, (ii) for weak 
tunnel couplings, and (iii) for large BCS gap.  
The  transition between these regions is studied numerically.
\end{abstract}


\maketitle

Several experimental groups have recently started to study the 
Josephson effect in ultra-small nanostructures,\cite{nazarov} where the 
supercurrent can be tuned via the gate voltage dependence of the
electronic levels of the nanostructure.  An important 
system class where supercurrents have been successfully observed 
\cite{tubeexp} is provided by carbon nanotube (CNT) quantum dots.  
In many cases, the experimental results compare quite well to 
predictions based on modeling the CNT dot as a spin-degenerate electronic level 
with $SU(2)$ spin symmetry, where the presence of
a repulsive on-dot charging energy $U$  may allow for a 
(normal-state) Kondo effect.  Depending on the ratio
$T_K/\Delta$, where $\Delta$ is the energy gap in the superconducting
electrodes and $T_K$ the Kondo temperature, 
theory\cite{glazman,rozhkov,slaveboson,su2numeric,largedelta,vecino} 
predicts a transition between a unitary (maximum) Josephson current
for $\Delta\ll T_K$, possible thanks to the survival of
the Kondo resonance in that limit, and a $\pi$-junction regime 
for $\Delta\gg T_K$, where the critical current is small and negative,
 i.e., the junction free energy $F(\varphi)$ has a minimum at phase 
difference $\varphi=\pi$ as opposed to the more common $0$-junction behavior.  

Recent progress has paved the way for the fabrication of very clean
CNTs, resulting in a new generation of quantum
transport experiments and thereby revealing interesting physics, e.g., 
spin-orbit coupling effects\cite{mceuen} or incipient Wigner 
crystal behavior.\cite{bockrath} In ultra-clean CNTs, 
the orbital degree of freedom ($\alpha=\pm$) reflecting clockwise and 
anti-clockwise motion around the CNT circumference  
(i.e., the two $K$ points) is
approximately conserved when electrons enter or leave the dot.\cite{aguado}
Due to the combined presence of this orbital ``pseudo-spin'' 
(denoted in the following by $T$) and the true electronic spin ($S$), 
an enlarged $SU(4)$ symmetry group can be realized.
In addition, a purely orbital $SU(2)$ symmetry arises 
when a Zeeman field is applied.  
Experimental support for this scenario has already been published 
\cite{su4exp} (for the case of semiconductor dots, 
see Ref.~\onlinecite{sasaki}), and several aspects have been
 addressed theoretically.\cite{aguado,su4theory} 
In particular, the $SU(4)$ Kondo regime is characterized by an enhanced Kondo 
temperature and exotic local Fermi liquid behavior,
where the Kondo resonance is asymmetric with respect to the 
Fermi level.  However, so far both experiment and theory have only studied
the case of normal-conducting leads, where conventional linear
response transport measurements cannot reliably distinguish the $SU(4)$
from the $SU(2)$ scenario.\cite{su4theory}
Here we provide the first theoretical study of the Josephson effect
for interacting quantum dots with (approximate) $SU(4)$ symmetry,
and find drastic differences compared to the standard $SU(2)$ picture.
In the Kondo limit, a qualitatively different current-phase relation (CPR)
is found, with the critical current smaller by a factor $\approx 0.59$.
The usual $\pi$-junction behavior is largely suppressed, but new
phases do appear and time-reversal symmetry can be spontaneously
broken.  Our predictions can be tested using  
state-of-the-art experimental setups, and offer clear signatures of 
the $SU(4)$ symmetry in very clean CNT quantum dots.  

\textit{Model and formal solution.---}
We study a quantum dot ($H_d$) contacted via a standard tunneling
Hamiltonian ($H_t$) to two identical superconducting 
electrodes ($H_{L/R}$),  $H=H_d+H_t+H_L+H_R$.
We assume that the dot has a spin- and orbital-degenerate
electronic level $\epsilon_{\alpha\sigma}= \epsilon$ with 
identical intra- and inter-orbital charging energy $U$,\cite{foot1}
$H_d = \epsilon \hat n + U \hat n(\hat n-1) /2$  with $\hat
n=\sum_{\alpha\sigma} d^\dagger_{\alpha \sigma} d^{}_{\alpha\sigma}$,
where $d^\dagger_{\alpha\sigma}$ creates a dot electron with spin
 $\sigma=\uparrow,\downarrow=\pm$ and orbital pseudo-spin 
projection $\alpha$.  Since the $\alpha=\pm$ states are related by
time-reversal symmetry (clockwise and anti-clockwise states are exchanged), 
we take the lead Hamiltonian as
\[
H_{j}  =  \sum_{{\bm k}\alpha\sigma} \xi_{\bm k} \
c_{j{\bm k}\alpha\sigma}^\dagger c_{j{\bm k}\alpha\sigma}^{} 
+  \sum_{{\bm k} \alpha} \left( \Delta e^{\mp i \frac{\varphi}{2}} \
c^\dagger_{j{\bm k}\alpha\uparrow} c^\dagger_{j,-{\bm k},-\alpha, \downarrow} 
+ {\rm h.c.} \right),
\]
where $c^\dagger_{j{\bm k}\alpha\sigma}$ creates an electron with
wavevector ${\bm k}$ in lead $j=L/R$, and $\xi_{\bm k}$ is 
the single-particle energy.  The tunneling Hamiltonian is 
$H_t = \sum_{j{\bm k}\sigma,\alpha\alpha'} \left(t\delta_{\alpha\alpha'}
+ \tilde t \delta_{\alpha,-\alpha'}\right) c^\dagger_{j{\bm k}\alpha\sigma}
d^{}_{\alpha'\sigma} + {\rm h.c.},$ 
where $t$ ($\tilde t$) describes orbital (non-)conserving tunneling
processes.  Following standard steps,\cite{rozhkov} 
the noninteracting lead fermions can now be integrated out. 
The partition function $Z(\varphi)=e^{-\beta F(\varphi)}$ at inverse
temperature $\beta$ then reads (we often set $e=\hbar=1$)
\begin{equation}\label{partition}
Z(\varphi) = {\rm Tr}_{d} \left( e^{-\beta H_d} {\cal T} 
e^{-\int_0^\beta d\tau d\tau'\  D^\dagger(\tau) \Sigma(\tau-\tau')
D(\tau') } \right) ,
\end{equation}
where the trace extends over the dot Hilbert space,
${\cal T}$ denotes time ordering, and we use the Nambu bispinor
$D=(d^{}_{e\uparrow}, d^\dagger_{e\downarrow}, d^{}_{o\uparrow},
d^\dagger_{o\downarrow})$ with even/odd linear combinations of
 the orbital states, $d_{e\sigma} = (d_{+,\sigma}+ d_{-,\sigma})/\sqrt{2}$
and $d_{o\sigma}=\sigma (d_{+,\sigma}-d_{-,\sigma})/\sqrt{2}$.
In this basis, the self-energy $\Sigma(\tau)$ representing the BCS leads
is diagonal in orbital space.  With the orbital mixing angle 
$\theta=2\tan^{-1} (\tilde t/t)$ and the normal-state
density of states $\nu_0 = 2\sum_{\bm k} \delta(\xi_{\bm k})$,
the even/odd channels are characterized by the hybridization widths
$\Gamma_{\nu=e,o} =  (1\pm \sin\theta)\ \Gamma$ with
$\Gamma = \pi \nu_0 (t^2+{\tilde t}^2)$.
In what follows, we study the zero-temperature limit and 
assume the wide-band limit\cite{nazarov} for the leads.
The Fourier transformed self-energy 
is then expressed in terms of the $2\times 2$ Nambu matrices
$\Sigma_{\nu=e,o}(\omega) = \frac{\Gamma_\nu}{\sqrt{\omega^2+\Delta^2}}
\left( \begin{array}{cc} -i\omega & \Delta \cos\frac{\varphi}{2} \\
\Delta \cos\frac{\varphi}{2} & -i\omega\end{array} \right).$
The result (\ref{partition}) will now be examined in several
limits. We start with the strong-correlation limit $U\to \infty$,
and later address the case of finite $U$.  Note that Eq.~(\ref{partition})
for $\theta=0$ corresponds to the $SU(4)$ symmetric case 
while for $\theta=\pi/2$ there is only one conducting channel with
non-zero transmission which, under certain conditions, corresponds to 
the usual $SU(2)$ model.

\vspace*{0.5cm}
\begin{figure}[ht!]
\begin{minipage}[t]{8.0cm}
\epsfig{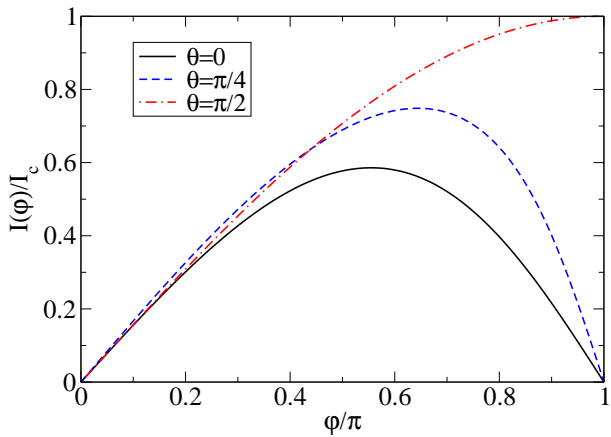}
\caption{\label{fig1} (Color online) 
Josephson CPR in the Kondo limit for various $\theta$. The $SU(4)$ case
corresponds to $\theta=0$, the $SU(2)$ case to $\theta=\pi/2$. 
The supercurrent is given in units of the unitary limit $I_c=e\Delta/\hbar$.
}
\end{minipage}
\end{figure}

\textit{Deep Kondo limit.---}
Let us first discuss the Kondo limit $T_K\gg \Delta$
in the quarter-filled case, $\epsilon<0$
and $\langle \hat n\rangle \approx 1$.  The Kondo temperature
is given by $T_K=D \exp(\pi\epsilon/4\Gamma)$\cite{aguado}
with bandwidth $D$.  As in the $SU(2)$ case,\cite{glazman}
the Josephson current at $T=0$ can be computed from 
local Fermi liquid theory, either using phase shift arguments or 
an equivalent mean-field slave-boson treatment.\cite{slaveboson}
The latter approach yields the self-consistent dot level $\tilde\epsilon$ and 
thereby the transmission probability for channel $\nu=e,o$,\cite{aguado}
\begin{equation}
{\cal T}_\nu = \frac{(1\pm \sin\theta)^2 T_K^2}{\tilde \epsilon^2 + (1\pm
\sin\theta)^2 T_K^2}, \quad \frac{\tilde \epsilon}{T_K} =
\frac{(1-\sin\theta)^{(\sin\theta+1)/4}}{(1+\sin\theta)^{(\sin\theta-1)/4}}.
\end{equation}
In the $SU(4)$ case ($\theta=0$), we have ${\cal T}_e={\cal T}_o=1/2$, while
the $SU(2)$ limit ($\theta=\pi/2$) has a decoupled odd channel, 
${\cal T}_e=1$ and ${\cal T}_o=0$.
The CPR covering the crossover from the $SU(4)$ to the $SU(2)$ Kondo regime
then follows as
\begin{equation}
I(\varphi) = \frac{e\Delta}{2\hbar} \sum_{\nu=e,o} \frac{{\cal T}_\nu 
\sin\varphi}{\sqrt{1-{\cal T}_\nu \sin^2\frac{\varphi}{2}}}.
\end{equation}
The known $SU(2)$ result\cite{glazman} is recovered
for $\theta=\pi/2$.  The $SU(4)$ CPR has a completely different shape,
as shown in Fig.~\ref{fig1}.
We note that the critical current $I_c={\rm max} [I(\varphi)]$ is suppressed 
by the factor $2-\sqrt{2}\approx 0.59$ relative to the
unitary limit $e\Delta/\hbar$ reached for the $SU(2)$ dot.  
The Josephson current in the deep Kondo 
regime is thus very sensitive to the $SU(4)$ vs $SU(2)$ symmetry.

\textit{Perturbation theory in $\Gamma$.---}
Next we address the opposite limit of very small $\Gamma\ll \Delta$, where
lowest-order perturbation theory in $\Gamma$ applies.  
After some algebra, Eq.~(\ref{partition}) for $\theta=0$
yields the CPR of a tunnel junction,
$I(\varphi)=I_c\sin(\varphi)$, where the critical current is
\begin{equation}\label{perturb}
I_c = \left( 4\Theta(\epsilon)-\Theta(-\epsilon) \right)
F(|\epsilon|/\Delta) \ I_0,
\end{equation}
 with the Heaviside function $\Theta$, the current scale
$I_0 = \Delta (\Gamma/\pi \Delta)^2$, and (see also Ref.~\onlinecite{novotny})
\[
F(x) = \frac{(\pi/2)^2 (1-x) - {\rm arccos}^2 x}{2x (1-x^2)}.
\]
In this $U\to \infty$ limit, the dot contains one electron for 
(finite) $\epsilon<0$, and thus we have spin $S=1/2$. 
Equation (\ref{perturb}) shows that such a magnetic
junction displays a $\pi$-phase. For the $SU(4)$ case, the ratio 
$I_c(-|\epsilon|)/I_c(|\epsilon|)=-1/4$ is twice smaller
than in the $SU(2)$ case, i.e., $\pi$-junction
behavior tends to be suppressed.  This tendency is
also confirmed for $U\ll \Delta$ (see below), where 
the $\pi$-phase is in fact essentially absent. The factor $1/4$ can
be understood in simple terms by counting the number of possible
processes leading to a Cooper-pair transfer through the 
dot.\cite{spivak,shimizu}.
When $\epsilon>0$, there are four possibilities corresponding to the
quantum numbers $(\alpha,\sigma)$ of the first electron entering the dot. 
However, for $\epsilon<0$ there is only one possibility since an
electron already occupies the dot, and then only one specific choice of 
$(\alpha,\sigma)$ allows for Cooper pair tunneling.  
This argument is readily generalized to the $SU(2N)$ case, where
the above ratio of critical currents is obtained as $-1/2N$.

\vspace*{0.6cm}
\begin{figure}[ht!]
\begin{minipage}[t]{8.0cm}
\epsfig{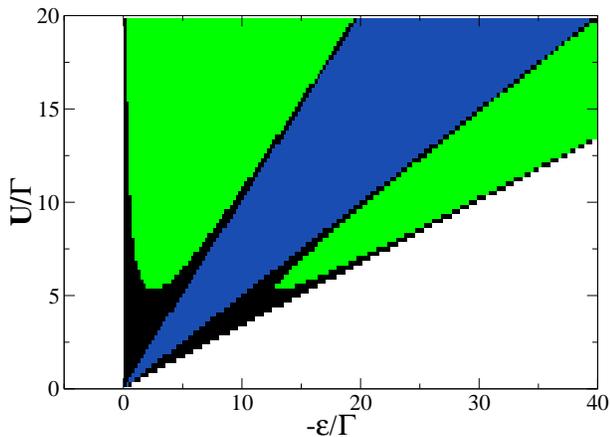}
\caption{\label{fig2} (Color online) 
Phase diagram for $\Delta\to \infty$.  White regions correspond to
 $(S,T)=0$, and green regions to $(S,T)=1/2$.
In the black regions, the ground state has $(S,T)=0$ 
for $\varphi=0$ and $(S,T)=1/2$ for $\varphi=\pi$. 
For the blue region, we have $(S,T)=0$ at $\varphi=0$ 
and $(S,T)=1$ at $\varphi=\pi$.\cite{footn} }
\end{minipage}
\end{figure}

\textit{Effective Hamiltonian for $\Delta\to \infty$.---}
The partition function (\ref{partition}) simplifies considerably
when $\Delta$ exceeds all other energy scales of interest. 
Then the dynamics is always confined to the subgap region (Andreev states),
and quasiparticle tunneling processes from the leads (continuum states)
are negligible.  In particular, this allows to study the case $U\ll \Delta$.
In fact, for $\Delta\to \infty$, with the Cooper pair 
operators $b^\dagger_1 = d^\dagger_{e\uparrow} d^\dagger_{o\downarrow}$ and
$b^\dagger_2 = d^\dagger_{o\uparrow} d^\dagger_{e\downarrow}$,
Eq.~(\ref{partition}) is equivalently described 
by the effective dot Hamiltonian 
\begin{equation}\label{delinf}
H_\infty = H_d + \cos(\varphi/2) \left [ \Gamma_e
 b_1 + \Gamma_o b_2 + {\rm h.c.} \right ].
\end{equation}
The resulting Hilbert space can be decomposed into three decoupled 
sectors\cite{footnot2}
according to spin $S$ and orbital pseudo-spin $T$ 
(notice that these quantities are localized on the dot for $\Delta\to \infty$).
The ground-state energy $E_g(\varphi)= {\rm min}(E_{(S,T)})$  then
determines the Josephson current $I(\varphi)=2\partial_\varphi E_g(\varphi)$. 
(i) The $(S,T)=0$ sector is spanned by the four 
states $\{ |0\rangle, b_1^\dagger|0\rangle,
b_2^\dagger|0\rangle, b_1^\dagger b_2^\dagger|0\rangle\}$, 
where $|0\rangle$ is the empty dot state.  The matrix representation 
reads
\[
H_{(S,T)=0} = \left( \begin{array}{cccc} 0 & \Gamma_e \cos\frac{\varphi}{2} & 
\Gamma_o \cos\frac{\varphi}{2} & 0 \\
\Gamma_e \cos\frac{\varphi}{2} & E_2 & 0 & \Gamma_o\cos\frac{\varphi}{2}\\
\Gamma_o \cos\frac{\varphi}{2} & 0 & E_2  & \Gamma_e\cos\frac{\varphi}{2}\\
0 & \Gamma_o \cos\frac{\varphi}{2} & \Gamma_e\cos\frac{\varphi}{2}
& E_4 \end{array} \right),
\]
with the eigenenergies $E_n= \epsilon n + U n(n-1)/2$ of the decoupled dot.
The lowest energy $E_{(S,T)=0}=E_2+z$ then follows from the smallest root
of the quartic equation
$\prod_\pm \left( z^2 - 2 z U - (\Gamma_e\pm \Gamma_o)^2 
\cos^2\frac{\varphi}{2} \right) = (E_4 z/2)^2.$
(ii) The $(S,T)=1/2$ sector can be decomposed into four
subspaces with one or three electrons according to $S_z=\pm 1/2$ and 
Cooper pair channel $\nu=e,o$.  The Hamiltonian is
$H^{(\nu)}_{(S,T)=1/2}=\left( \begin{array}{cc} E_1 & 
\Gamma_\nu\cos\frac{\varphi}{2}
\\ \Gamma_\nu \cos\frac{\varphi}{2} & E_3 \end{array}\right)$, 
where $H^{(e)}_{(S,T)=1/2}$ operates in the subspace
spanned by $\{ d^\dagger_{o\uparrow}|0\rangle, b_1^\dagger 
d^\dagger_{o\uparrow}| 0\rangle \}$ for $S_z=+1/2$, and
$\{ d^\dagger_{e\downarrow}|0\rangle, b_1^\dagger 
d^\dagger_{e\downarrow}| 0\rangle \}$ for $S_z=-1/2$.
(Similarly, the subspaces corresponding to $H^{(o)}_{(S,T)=1/2}$ 
are obtained by letting $d^\dagger_{\nu\sigma}\to d^\dagger_{\nu,-\sigma}$
and $b_1^\dagger\to b_2^\dagger$.)
With $\Gamma_e\ge \Gamma_o$, the lowest energy is
$E_{(S,T)=1/2}=  \left( E_1+E_3 - \left[(E_3-E_1)^2 + 4\Gamma_e^2 \cos^2
(\varphi/2)\right]^{1/2} \right)/2.$
(iii) Finally, the $(S,T)=(1,0)$ sector is spanned by the two 
uncoupled two-particle states $d^{\dagger}_{e,\sigma}
d^\dagger_{o,\sigma}|0\rangle$, with $\varphi$-independent 
energy $E_{S=1,T=0}=E_2$. In addition, there are two decoupled $(S,T)=(0,1)$ 
states $d_{\nu\uparrow}^\dagger d_{\nu\downarrow}^\dagger|0\rangle$ 
with the same energy $E_2$. In the limit $\Delta\to \infty$,
this $(S,T)=1$ sector is energetically unfavorable except 
possibly at $\varphi=\pi$.

\vspace*{0.6cm}
\begin{figure}[ht!]
\begin{minipage}[t]{8.0cm}
\epsfig{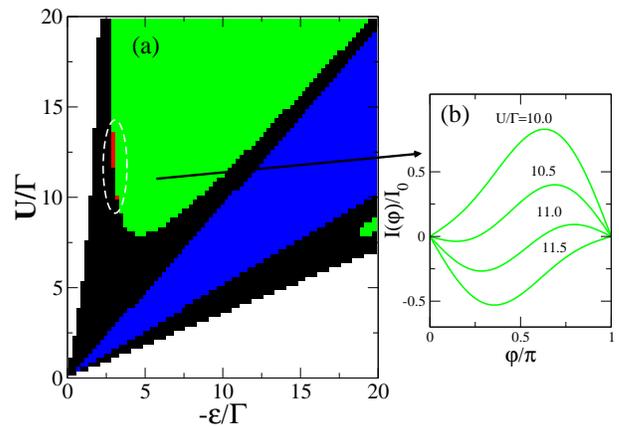}
\caption{\label{fig3} (Color online) 
(a) Same as Fig.~\ref{fig2} but for $\Delta=10\Gamma$
within the zero-bandwidth limit for the leads (see text).
Although the $\Delta\to\infty$ phase diagram is basically reproduced,
for finite $\Delta$, the $(S,T)=1/2$ phase (green) exhibits a crossover from 
0- to $\pi$-junction behavior for $U \simeq \Delta$, as illustrated
in panel (b), where the CPR is shown for $\epsilon/\Gamma=-5$
and several $U$; the current is normalized to $I_0$, 
see Eq.~(\ref{perturb}).  Moreover, a phase with $(S,T)=0$ at 
$\varphi=0$ and $(S,T)=1/2$ at $\varphi=\pi$ 
appears, where (contrary to the ``black'' phase) $\varphi=\pi$
corresponds to the lowest energy ($\pi'$-behavior), 
indicated in red [within the dashed ellipses in panel (a)].} 
\end{minipage}
\end{figure}

\textit{Phase diagram for $\Delta\gg \Gamma$.---}
Next we discuss the resulting phase diagram in the $SU(4)$ limit ($\theta=0$).
The result for $\Delta\to \infty$ 
is shown in Fig.~\ref{fig2} in the $U-\epsilon$ plane. 
The phases are classified according to the three sectors defined 
above.\cite{footn} The reported phases are specific for the $SU(4)$ symmetry
and are qualitatively different from the standard $SU(2)$ case.
We observe that the $\varphi$-dependence of the $\Delta\to \infty$ 
ground-state energy implies $0$-junction behavior for both $S=0$ and $S=1/2$.
While the magnetic $S=1/2$ sector often represents a 
$\pi$-junction,\cite{glazman,rozhkov,su2numeric} in 
multi-level dots there is no direct connection between the 
spin and the sign of the Josephson coupling.\cite{shimizu}
The $\pi$-phase found under perturbation theory [Eq.~(\ref{perturb}) for
$\epsilon<0$] is in fact restricted to 
the regime $U\gg \Delta$, while for $U\ll \Delta$,
the $S=1/2$ state displays a $0$-phase.  In the intermediate 
regime one should therefore observe a crossover between those
two behaviors.  Interestingly, there are parameter regions with 
a spin/pseudo-spin transition as $\varphi$ varies.
For instance, the ``black'' regions in Fig.~\ref{fig2} correspond to a 
mixed state with $(S,T)=0$ at $\varphi=0$ and $(S,T)=1/2$ at $\varphi=\pi$,
while for the ``blue'' region, the ground state is in the $(S,T)=0$ 
sector except at $\varphi=\pi$ where it crosses to the $(S,T)=1$ sector.

We find that these phases are also observable at finite 
$\Delta \agt \Gamma$, where we have employed two complementary approaches.  
First, a full numerical solution is possible when approximating each 
electrode by a single site (zero-bandwidth limit),
which can provide a satisfactory, albeit not
quantitative, understanding of the phase diagram.\cite{vecino}
Second, one can go beyond the above $\Delta\to \infty$ limit by
including cotunneling processes in a systematic way.
Both approaches give essentially the same results, and here we 
only show results from the single-site model. 
As can be observed in Fig.~\ref{fig3}(a), the overall features of the
$\Delta\to\infty$ phase diagram are reproduced for finite $\Delta$, 
with somewhat shifted boundaries between the different regions.  
In particular, in the ``green'' ($S=1/2$) regime, 
this calculation captures the mentioned transition from a $0$-junction 
at $\Delta\gg U$ to a $\pi$-junction at $\Delta\ll U$, 
as illustrated in Fig.~\ref{fig3}(b).
Consequently, for finite $\Delta$, the ``black'' phase may now
have lowest energy at $\varphi=\pi$, implying 
the $\pi'$-phase \cite{rozhkov,su2numeric,vecino}
indicated in ``red'' in Fig.~\ref{fig3}(a).  
%We stress that in the $SU(4)$ case and within this range of parameters, 
%this $0-\pi$ transition is a perfectly smooth crossover, while in the 
%$SU(2)$ case, one has necessarily sharp quantum phase transitions.
Finally, for the junctions with $U/\Gamma=10.5$ and $11$ 
in Fig.~\ref{fig3}(b), the ground state is realized at phase difference 
$0<\varphi<\pi$, which implies that time-reversal symmetry is
spontaneously broken here.

To conclude, we have studied the Josephson current in $SU(4)$ symmetric
quantum dots, including the crossover to the standard $SU(2)$ symmetric
case.  Contrary to normal-state transport, the supercurrent is very
sensitive to the symmetry group, and should allow to observe clear
signatures of the $SU(4)$ state in ultra-clean CNT dots.
In particular, the $\pi$-phase is largely suppressed, the CPR in the
Kondo limit has a distinctly different shape and a smaller 
critical current, and the phase diagram turns out to be quite rich.
In addition, following Ref.~\onlinecite{noise}, we expect a strongly reduced 
thermal noise in the deep $SU(4)$ Kondo regime since (in contrast
to the $SU(2)$ case) there are two channels with
imperfect transmission. Future theoretical work is  needed
to give a quantitative understanding of the crossover between the various
regimes discussed above.  
%Numerical renormalization group
%and quantum Monte Carlo calculations should be particularly
%useful in that regard.  

This work was supported by the SFB TR/12 of the DFG, the 
EU network HYSWITCH, the ESF network INSTANS and by the Spanish
MICINN under contracts FIS2005-06255 and FIS2008-04209.


\begin{thebibliography}{99}

\bibitem{nazarov}
Yu.V. Nazarov and Ya.M. Blanter, \textit{Quantum transport: Introduction to
nanoscience} (Cambridge University Press, 2009).

\bibitem{tubeexp}
A.Y. Kasumov \textit{et al.}, Science {\bf 284}, 1508 (1999);
A.F. Morpurgo \textit{et al.}, \textit{ibid.} {\bf 286}, 263 (1999);
M.R. Buitelaar, T. Nussbaumer, and C. Sch\"onenberger,
Phys. Rev. Lett. {\bf 89}, 256801 (2002);
P. Jarillo-Herrero, J.A. van Dam, and L.P. Kouwenhoven,
Nature {\bf 439}, 953 (2006);
J.-P. Cleuziou \textit{et al.}, Nat. Nanotechnol. {\bf 1}, 53 (2006);
H.I. Jorgensen \textit{et al.}, Phys. Rev. Lett. {\bf 96}, 207003 (2006);
A. Eichler \textit{et al.}, Phys. Rev. B {\bf 79}, 161407(R) (2009). 

\bibitem{glazman}
L.I. Glazman and K.A. Matveev, JETP Lett. {\bf 49}, 659 (1989).


\bibitem{rozhkov}
A.V. Rozhkov and D.P. Arovas, Phys. Rev. Lett. {\bf 82}, 2788 (1999).

\bibitem{slaveboson}
A.A. Clerk and V. Ambegaokar, Phys. Rev. B {\bf 61}, 9109 (2000);
A.V. Rozhkov and D.P. Arovas, \textit{ibid.} {\bf 62}, 6687 (2000).

\bibitem{su2numeric}
F. Siano and R. Egger, Phys. Rev. Lett. {\bf 93}, 047002 (2004);
M.S. Choi \textit{et al.}, Phys. Rev. B {\bf 70}, 020502(R) (2004); 
G. Sellier \textit{et al.}, \textit{ibid.} {\bf 72}, 174502 (2005);
C. Karrasch, A. Oguri, and V. Meden, \textit{ibid.} {\bf 77}, 024517 (2008);
M. Governale, M.G. Pala, and J. K\"onig, \textit{ibid.} {\bf 77}, 134513 (2008).

\bibitem{largedelta}
A. Zazunov, A. Schulz, and R. Egger, Phys. Rev. Lett. {\bf 102}, 047002 (2009);
T. Meng, S. Florens, and P. Simon,  Phys. Rev. B {\bf 79}, 224521 (2009).

\bibitem{vecino}
E. Vecino, A. Mart{\'i}n-Rodero, and A. Levy Yeyati, Phys. Rev. B
{\bf 68}, 035105 (2003).

\bibitem{mceuen}
F. Kuemmeth \textit{et al.}, Nature {\bf 452}, 448 (2008).

\bibitem{bockrath}
V.V. Deshpande and M. Bockrath, Nat. Phys. {\bf 4}, 314 (2008);
V.V. Deshpande \textit{et al.}, Science {\bf 323}, 106 (2009).

\bibitem{aguado}
J.S. Lim \textit{et al.}, Phys. Rev. B {\bf 74}, 205119 (2006). 

\bibitem{su4exp}
P. Jarillo-Herrero \textit{et al.}, Nature {\bf 434}, 484 (2005);
A. Makarovski \textit{et al.}, Phys. Rev. B {\bf 75}, 241407(R) (2007);
A. Makarovski, J. Liu, and G. Finkelstein, Phys. Rev. Lett.
{\bf 99}, 066801 (2007); T. Delattre \textit{et al.}, 
Nat. Phys. {\bf 5}, 208 (2009). 

\bibitem{sasaki}
S. Sasaki \textit{et al.}, Phys. Rev. Lett. {\bf 93}, 017205 (2004).

\bibitem{su4theory} 
L. Borda \textit{et al.}, Phys. Rev. Lett. {\bf 90}, 026602 (2003); 
M.S. Choi, R. L{\'o}pez, and R. Aguado, \textit{ibid.} {\bf 95}, 067204 (2005); 
K. Le Hur, P. Simon, and D. Loss, Phys. Rev. B {\bf 75}, 035332 (2007);
C.A. B\"usser and G.B. Martins, \textit{ibid.} {\bf 75}, 045406 (2007);
C. Mora, X. Leyronas, and N. Regnault, Phys. Rev. Lett. 
{\bf 100}, 036604 (2008); P. Vitushinsky, A.A. Clerk, and 
K. Le Hur, \textit{ibid.} {\bf 100}, 036603 (2008); 
F.B. Anders \textit{et al.}, \textit{ibid.} {\bf 100}, 086809 (2008).


\bibitem{foot1}
It is straightforward to allow for orbital or Zeeman fields, 
or for more general interactions. 
We also consider identical tunnel couplings
between the dot and both electrodes.  Asymmetries 
produce similar effects as the orbital mixing
 $\tilde t$ in $H_t$.\cite{aguado} 

\bibitem{novotny}
T. Novotn{\' y}, A. Rossini, and K. Flensberg, Phys.
Rev. B \textbf{72}, 224502 (2005).

\bibitem{spivak}
B.I. Spivak and S.A. Kivelson, Phys. Rev. B {\bf 43}, 3740 (1991).

\bibitem{shimizu}
Y. Shimizu, H. Horii, Y. Takane, and Y. Isawa, J. Phys. Soc. Jpn.
{\bf 5}, 1525 (1998); A.V. Rozhkov, D.P. Arovas, and F. Guinea,
Phys. Rev. B {\bf 64}, 233301 (2001).

\bibitem{footnot2}
Different phases can be labeled by $S^2 + T^2$, and we use the notation
$(S,T)=0$ for $S=T=0$,  $(S,T)=1/2$ for $S=T=1/2$,
and $(S,T)=1$ for both $(S,T)=(1,0)$ and $(0,1)$.

\bibitem{footn}
Strictly speaking, precisely at $\Delta=\infty$ there are degeneracies
that make the classification ambiguous. However,  at finite $\Delta$,
Fig.~\ref{fig3} indicates that the reported phases are stable. 
We have also confirmed their stability for $\theta\ne 0$.

\bibitem{noise}
A. Mart{\'i}n-Rodero, A. Levy Yeyati, and F.J. Garc{\'i}a-Vidal,
Phys. Rev. B {\bf 53}, R8891 (1996).

\end{thebibliography}
\end{document}